# Designer spin order in diradical nanographenes

Yuqiang Zheng[1†], Can Li[1†], Chengyang Xu[1], Doreen Beyer[2], Xinlei Yue[1], Yan Zhao[1], Guanyong Wang[1], Dandan Guan[1,3], Yaoyi Li[1,3], Hao Zheng[1,3], Canhua Liu[1,3], Junzhi Liu[4], Xiaoqun Wang[1,3], Weidong Luo[1,5], Xinliang Feng[2]*, Shiyong Wang[1,3]*, Jinfeng Jia[1,3]*

[1]*Key Laboratory of Artificial Structures and Quantum Control (Ministry of Education), Shenyang National Laboratory for Materials Science, School of Physics and Astronomy, Shanghai Jiao Tong University, Shanghai 200240, China.*

[2]*Center for Advancing Electronics Dresden & Department of Chemistry and Food Chemistry, Technische Universität Dresden, 01062 Dresden, Germany.*

[3] *Tsung-Dao Lee Institute, Shanghai Jiao Tong University, Shanghai, 200240, China.*

[4] *Department of Chemistry and State Key Laboratory of Synthetic Chemistry, The University of Hong Kong, Pokfulam Road, Hong Kong, China.*

[5]*Institute of Natural Sciences, Shanghai Jiao Tong University, Shanghai, 200240, China.*

[†]These authors contributed equally to this work.

*Corresponding Authors: xinliang.feng@tu-dresden.de, shiyong.wang@sjtu.edu.cn, jfjia@sjtu.edu.cn

## Abstract

The magnetic properties of carbon materials are at present the focus of an intense research effort in physics, chemistry and materials science due to their potential applications in spintronics and quantum computations. Although the presence of spins in open-shell nanographenes has been recently confirmed, the ability to control magnetic coupling sign has remained elusive, but the most desirable. Here, we demonstrate an effective approach of engineering magnetic ground states in atomically precise open-shell bipartite/nonbipartite nanographenes using combined scanning probe techniques and mean-field Hubbard model calculations. The magnetic coupling sign between two spins has been controlled via breaking bipartite lattice symmetry of nanographenes. In addition, the exchange-interaction strength between two spins has been widely tuned by finely tailoring their spin density overlap, realizing a large exchange-interaction strength of 42 meV. Our demonstrated method provides ample opportunities for designer above-room-temperature magnetic phases and functionalities in graphene nanomaterials.

**Introduction**

Nanographenes (NGs) with certain π-electron topologies are predicated to host manifold magnetic properties, promising for designer miniaturized spin-based devices and for quantum computations[1–15]. The study of magnetic nature of NGs traces back to 1950s, where Clar established an empirical method to determine the number of unpaired electrons in NGs by drawing non-kekulé structure[16]. A typical example is the bowtie-shaped NG, also called Clar goblet, which holds two unpaired electrons with an antiferromagnetic coupled ground state[8,17]. In 1989, Lieb came up with a theorem for bipartite lattice stating that NGs with sublattice imbalance host a net spin of $S=(N_A-N_B)/2$ with $N_A/N_B$ denoting the number of carbon atoms in each triangular sublattice $A/B$ of graphene honeycomb lattice[3]. Very recently, Ortiz et al studied the exchange interactions of bipartite NGs by using a set of combined theoretical methods at different levels to determine both the ground state and the first few excited states[18]. They found that magnetic exchange of diradical bipartite NGs depends crucially on sublattice imbalance, point symmetry group, and wavefunction overlap of zero modes.

Departing from the above bipartite NGs, certain NGs with non-hexagonal rings, also called nonbipartite NGs, host nontrivial magnetic features and have seldomly been addressed. Previous theory calculations demonstrated that incorporation of carbon tetragons/pentagons in zigzag graphene nanoribbons switches the spin orientation at each edge, and such spin switches can lift the degeneracy between the two spin propagation channels[13,14]. Since magnetic NGs are usually unstable and reactive due to the presence of radicals, the fabrication of unsubstituted NGs has remained elusive for a long time, although the addition of substituents has allowed the solution synthesis of few open-shell NG cores and verification of their magnetic ground state via electron paramagnetic resonance measurements[19]. Recent advances in on-surface chemistry have made it possible to fabricate reactive graphene nanostructures with atomic precision and unprecedented tunability by using small molecules as building blocks, with examples of Clar goblet[17], zigzag graphene nanoribbons[20,21], Chiral graphene nanoribbons[22,23] and NGs with sublattice imbalance[24–29]. As demonsrated by Li et al, the incorporation of pentagon rings in chiral graphene nanoribbons introduces one unpaired electron and two nearby spins can antiferromagnetic coupled together[23]. We recently demonstrated that nanographenes with such pentagon rings also host net spins and the coupling strength can be engineered by tailoring spin density overlap at the connecting region[30].

Here, we further demonstrate a generic approach to control spin orders inside diradical NGs. Incorporation of a pentagon into a magnetic diradical NG drives the ferromagnetic coupled ground state into antiferromagnetic coupled ground state. The nature behind this exchange interaction, as we have revealed, is a local reversal of spin density sign at the connecting region of the dimer via breaking the bipartite lattice symmetry of NGs. This exchange mechanism is expected to be generic and widespread, because it relies solely on very general features of the nonbipartite character of NGs. Moreover, since it is feasible to precisely tailor the chemical structure of NGs by using on-surface synthesis, the established approach can be extended to design any artificial quantum spin systems, opening wide possibility for fundamental research and applications in spintronics and quantum information technologies.

**Results**

**Mean-field Hubbard model calculations on diradical nanographenes.** Figure 1 provides a paradigm of such exchange mechanism. The bipartite single radical NG considered (cf. Fig. 1a) hosts 14 carbon atoms in *A* sublattice and 15 carbon atoms in *B* sublattice, as marked by hollowed and solid spheres, respectively. As per Lieb's theorem, this NG should host a single spin of $S=1/2$ due to sublattice imbalance. Mean-field Hubbard model and spin-polarized density functional theory (SP-DFT) calculations also confirm the presence of one unpaired electron with enhanced spin density distribution at the right side (cf. Fig. 1b and supplementary Fig. 3-4). Figure 1d depicts a nonbipartite NG, which is similar as the above bipartite NG but with a single C-C bond, as marked by the red line, connecting two carbon atoms belonging to the same sublattice. The addition of this bond introduces a pentagon ring into the NG, and breaks the bipartite lattice symmetry. We refer to the bipartite NG and the nonbipartite NG as BNG and nBNG hereinafter, respectively. The nBNG also hosts a single spin of $S=1/2$ as confirmed by SP-DFT and Hubbard model calculations (cf. Fig. 1g). The overall spin density distribution of nBNG is similar as that of BNG with enhanced spin density at the right side. However, the spin density distribution at the left terminal phenyl ring of nBNG, as marked by dashed squares, is drastically different from that of BNG, where the sign of spin density gets reversed at this phenyl group as a result of the presence of the red C-C bond.

To reveal the effects of spin density sign reversal in the nBNG on the spin exchange interaction, the ground states of different NG dimer configurations have been determined. Figure 1c depicts a BNG-BNG dimer with 28 carbon atoms in $A$ sublattice and 30 carbon atoms in $B$ sublattice, hosting a ferromagnetic coupled ground state of $S=1$ as per Lieb's theorem. The two coupled spins have a magnetic exchange-interaction strength of 7 meV as obtained by Hubbard model calculations. In sharp contrast to the diradical BNG-BNG dimer, the BNG-nBNG dimer in Fig. 1f hosts an antiferromagnetic coupled ground state with $S=0$. We attribute the transition from ferromagnetic coupled ground state in the BNG-BNG dimer to antiferromagnetic coupled ground state in the BNG-nBNG dimer to the presence of on-site Coulomb repulsion. Since each carbon atom provides a $P_z$ orbital and one π electron, NGs are ideal two-dimensional half-filling systems. At half filling, spin up density prefers to populate at one sublattice while spin down density at another in order to minimize the on-site Coulomb repulsion (to prevent a double occupation of one $P_z$ orbital at the same carbon site)[3]. Since the spin density sign at the connecting region is reversed at the nBNG side in the dimer, the spin exchange sign will also get reversed to minimize the on-site Coulomb repulsion (Otherwise, the two adjacent carbon sites at the connecting region will host the same spin density sign, driving the system to a higher energy due to on-site Coulomb repulsion). Additional calculations have been made on other configurations of BNG-BNG dimers, nBNG-nBNG dimers, nBNG-BNG dimers and BNG-nBNG dimers (cf. supplementary Fig. 5). All the configurations agree with the previous physics picture: once the spin density sign gets reversed at the connecting region, the sign of magnetic exchange between the two spins will also get reversed. This unique property of nBNG provides a generic approach to control spin order in nBNGs.

**Kondo resonances of nanographenes with *S*=1/2.** We demonstrate the proposed magnetic exchange mechanism for diradical NGs experimentally. Fig. 2a and 2b depict the chemical structure and nc-AFM image of on-surface synthesized BNG monomers. The BNG hosts a methyl group which exhibits as a bright protrusion in nc-AFM imaging. As confirmed by our DFT calculations, the presence of methyl group does not affect the magnetic properties compared to the counterpart without a methyl group (cf. Supplementary Fig. 4). In addition, the magnetic nanographenes are embedded in a non-magnetic polymer (cf. Supplementary Fig. 1). As confirmed theoretically and experimentally, the presence of neighboring non-magnetic units does not affect the magnetic properties of embedded magnetic nanographenes (cf. Supplementary Fig. 6 and 8). The presence of a net spin of *S*=1/2 in the bipartite NG is confirmed by Kondo resonance, which gives a sharp peak at Fermi level in differential conductance (d*I*/d*V*) spectra. The Kondo resonance originates from the coupling of the magnetic impurity by Au(111) electron reservoir, and can be understood by single impurity Anderson Model[31,32]. This model gives a temperature dependent Kondo peak width with full width at half maximum (FWHM) of $\Gamma = \sqrt{(\alpha k_B T)^2 + (2k_B T_K)^2}$, where $T$ is the temperature, $T_K$ is the Kondo temperature, and $\alpha$ is the slope of linear growth of the width at $T \gg T_K$. The experimentally obtained temperature dependent FWHM in Fig. 2h can be well fitted by the previous equation, giving a Kondo temperature of 39 K. The Kondo peak is expected to response to an external magnetic field, and start to split for magnetic field up to $B_C \approx 0.5 K_B T_K/(g\mu_B)$ [32]. The threshold magnetic field for Kondo peak splitting in our case is estimated to be 15 T, which is beyond the capability (10 T) of our system (We notice the Kondo peak becoming broadened as increasing magnetic field, but we didn't observe Kondo peak splitting for magnetic field up to 9.5 T (cf. Fig. 2f)). The net single spin in nBNG monomers has also been carefully studied[30], and similar observations have been obtained (cf. Supplementary Fig. 7).

**Spin flip spectroscopy of diradical nanographenes.** The magnetic exchange for diradical NG dimers has been detected by spin flip spectroscopy, where inelastic tunneling electrons can excite the ground state into an excited state when the energy of tunneling electrons exceeds the energy difference between the two states[17,23,33–35]. The chemical structure and nc-AFM image of an achieved BNG-BNG dimer is depicted in Fig. 3a. In contrast to a sharp resonance obtained on the BNG monomer, d$I$/d$V$ spectra made on a BNG-BNG dimer exhibits three resonances, with a sharp peak at fermi level and two side shallowed shoulders at -7 meV and 7 meV, respectively. These features originate from the screening of a net spin of $S$=1 by Au(111) surface electrons (zero energy Kondo peak) and spin-flip excitations (two side peaks) from ground triplet state to excited singlet state. From the spin-flip feature, we can directly obtain the magnetic exchange-coupling strength of 7 meV, which agrees with Hubbard model calculations (7.7 meV). To qualitatively capture the scattering process inside an STM junction, we simulate our d$I$/d$V$ spectra using a perturbative approach up to third order using codes provided by Ternes[36], where the simulated spectra nicely reproduce the experimental features (cf. Supplementary Fig. 9). Fig. 3d depicts the chemical structure and nc-AFM image of a BNG-nBNG (from up to down) hetero-dimer. In sharp contrast to BNG-BNG dimer, d$I$/d$V$ spectra exhibit two symmetric steps at both positive and negative bias, indicating an antiferromagnetic ground state with $S=0$. The two steps at the same bias amplitude are due to spin excitation from the ground singlet state to excited triplet state, giving an exchange-coupling strength of 6 meV. Except for the spectra features, constant-height Kondo maps, reflecting the spin density distribution, also nicely agree with Hubbard model calculations (cf. Fig. 3b and 3e), further confirming our elucidation of d$I$/d$V$ spectra. These observations prove that breaking bipartite lattice symmetry will reverse the magnetic order between the two spins in diradical nBNGs. We also examined other dimer configurations, in support of the proposed exchange mechanism (cf. Supplementary Fig. 6).

**Engineering exchange-coupling strength between two spins.** The magnetic exchange-coupling strength can also be widely engineered in diradical NG dimers by tuning wavefunction (spin density) overlap of the zero-energy states[18]. The exchange-coupling strength can be expressed as $J = 2U \sum_i |\phi_1(i)|^2 |\phi_2(i)|^2$, where $U$ denotes the on-site Coulomb repulsion and $\phi_1/\phi_2$ the wavefunction of the zero-energy state of up/down unit in the diradical dimer, respectively. Figure 4a shows a BNG-BNG dimer with minimal spin density overlap. d$I$/d$V$ spectra taken on both BNG units exhibit a sharp Kondo resonance, similar to the features obtained on individual BNG monomers. This suggests that the two spins in this dimer has negligible exchange interaction due to negligible spin density overlap. We also examined another BNG-BNG dimer with maximum spin density overlap (cf. Fig. 4b). d$I$/d$V$ spectra reveal two steps located at -42 mV and 42 mV, suggesting an exchange-coupling strength of 42 meV (cf. Fig. 4d and 4e). The exchange-coupling interaction of different dimer configurations qualitatively agree with Hubbard model calculations (cf. Fig. 4f). These results demonstrate the wide tunability of magnetic exchange interaction in NGs, with implications for spintronic devices working above room temperature.

**Discussion**

We demonstrate an exotic approach of controlling magnetic exchange sign and strength in atomically precise diradical NG dimers by using combined non-contact atomic force microscopy, scanning tunneling microscopy/spectroscopy as well as model calculations. The magnetic coupling sign transition is due to the spin density sign reversal at the connecting region by breaking the bipartite lattice symmetry of nanographene. As summarized in Table 1, we have calculated and checked all the dimer configurations (cf. details in supplementary Fig. 5 and 6). Due to the negligible spin density overlap, all dimers in the *C1* configuration host two spins with negligible magnetic coupling strength, which can be treated as two uncoupled spins at a finite temperature. In the *C2* configuration, due to the revseral of spin density sign at the connecting region, both BNG-BNG and nBNG-BNG dimers host the ground state of $S=1$, while both nBNG-nBNG and BNG-nBNG dimers host the ground state of $S=0$. In the *C3* configuration, due to the strong spin density overlap, large magnetic exchange coupling up to 40 meV has been achieved.

The nature behind this exchange mechanism is the minimization of on-site Coulomb repulsion in half-filling NGs. We demonstrated, by tailoring the spin density sign and spatial overlap, the magnetic coupling sign and strength can be widely tuned. Our method reported herein provides ample opportunities for designer above-room-temperature magnetic phases and functionalities in graphene nanomaterials, and could be extended to study many-body effects in low-dimensional π-electron quantum spin systems.

**Methods**

Sample preparation and characterization were carried out using a commercial low-temperature Unisoku Joule-Thomson scanning probe microscopy under ultra-high vacuum conditions ($3\times10^{-10}$ mbar). The high magnetic field spectroscopy measurements were carried out using a commercial Unisoku 1300 system. Au(111) single-crystal was cleaned by cycles of argon ion sputtering, and subsequently annealed to 800K to get atomically flat terraces. Molecular precursors of 9,10-bis(4-bromo-2,6-dimethylphenyl)anthracene were thermally deposited on the clean Au(111) surface, and subsequently annealed to different temperatures to fabricate magnetic NGs (cf. Supplementary Fig. 1). Afterwards, the sample was transferred to a cryogenic scanner at 4.9K (1.1K) for characterization. Carbon monoxide molecules are dosed onto the cold sample around 7 K ($4\times10^{-9}$ mbar, 2 minutes). To achieve ultra-high spatial resolution, CO molecule is picked up from Au surface to the apex of tungsten tip. A quartz tuning fork with resonant frequency of 31 KHz has been used in nc-AFM measurements. A lock-in amplifier (589 Hz, 0.1-0.5 mV modulation) has been used to acquire dI/dV spectra. The spectra were taken at 4.9K unless otherwise stated.

The tight binding (TB) calculation of the STM images were carried out in the $C\ 2P_z$-orbital description by numerically solving the Mean-Field-Hubbard Hamiltonian with nearest-neighbor hopping:

$$\hat{H}_{MFH} = \sum_{<ij>,\sigma} -t_{ij} c_{i,\sigma}^+ c_{j,\sigma}^- + U\sum_{i,\sigma}\langle n_{i,\sigma}\rangle n_{i,\bar\sigma} - U\sum_i \langle n_{i,\uparrow}\rangle\langle n_{i,\downarrow}\rangle \tag{1}$$

with $t_{ij}$ is the nearest-neighbor hopping term depending on the bond length between C atoms and $c_{i,\sigma}^+$ and $c_{j,\sigma}^-$ denoting the spin selective ($\sigma=\{\uparrow,\downarrow\}$) creation and annihilation operators on the atomic site i and j, $U$ the on-site Hubbard parameter (with $U$=3.5eV), $n_{i,\sigma}$ the number operator and $\langle n_{i,\sigma}\rangle$ the mean occupation number at site i. A hopping integral of 2.7 eV is used in our calculation unless otherwise stated. Numerically solving the model Hamiltonian yields the energy Eigenvalues $E_i$ and the corresponding Eigenstates $\alpha_{i,j}$ (amplitude of state *i* on site *j*) from which the wave functions are computed assuming Slater type atomic orbitals:

$$\psi_i(\vec{r}) = \sum_j \alpha_{i,j} \cdot (z-z_j)\, exp(-\zeta|\vec{r}-\vec{r}_j|) \tag{2}$$

with $\zeta$=1.625 a.u. for the carbon $2P_z$ orbital. The charge density map ρ(x,y) for a given energy range [$\varepsilon_{min},\varepsilon_{max}$] and height $z_0$ is then obtained by summing up the squared wave functions in this chosen energy range.

$$\rho(x,y) = \sum_{i, \varepsilon_i \in [\varepsilon_{min}, \varepsilon_{max}]} \psi_i^2(x,y,z_0) \tag{3}$$

Constant charge density maps are taken as a first approximation to compare with experimental STM images.

Spin-polarized DFT calculations were performed by using local spin-density approximation (LSDA)[37] and generalized gradient approximation (GGA)[38] method. The first-principles calculations were carried out with the Vienna Ab initio Simulation Package (VASP)[39]. H atoms were included to saturate dangling bonds. We add 10 Å of vaccum in the direction perpendicular to the sheet and vacuum of 10 Å in two orthogonal directions in plane to eliminate the periodic effect of the system. A cutoff of 700 eV was used for convergence of the system for the short C-C bond and C-H bond. We optimized the geometric structure with the initial bond length of C-C bond 1.42 Å which was obtained from single sheet of graphene with LDA and GGA.

**Data availability**

The datasets generated and/or analysed during the current study are available from the corresponding author on reasonable request.

**Code availability**

The tight-binding calculations were performed using a custom-made code on the MATLAB platform. Details of this tight-binding code can be obtained from the corresponding author on reasonable request.

**Acknowledgements**

We acknowledge Prof. Markus Ternes for fruitful discussions. S. Wang acknowledge financial support from National Natural Science Foundation of China (No. 11874258, 12074247) and Fok Yin Tung foundation. J. Liu is grateful for the startup funding from the University of Hong Kong and the funding support from ITC to the SKL. This work is also supported by the Ministry of Science and Technology of China (Grants No. 2016YFA0301003, No. 2016YFA0300403), the National Natural Science Foundation of China (Grants No. 11521404, No. 11634009, No. 11574202, No. 11874256, No. 11790313, No. 11674226, No. U1632102, No. 11674222, No. U1632272 and No. 11861161003), and the Strategic Priority Research Program of Chinese Academy of Sciences (Grant No. XDB28000000).

**Author contributions**

S.W. conceived the experiments, S.W., J.J. and X.F. supervised the project, Y.Q.Z. and Y.Z. performed the STM/STS and nc-AFM experiments. C.L. performed the mean-field Hubbard model calculations. C.X. and W.L. performed the DFT calculations. D.B. J.L. and X.F. synthesized the molecular precursor. S.W., Y.Q.Z., and C.L. wrote the paper. All authors discussed the results and implications and commented on the manuscript at all stages.

**Competing interests**

The authors declare no competing interests.

# Figures

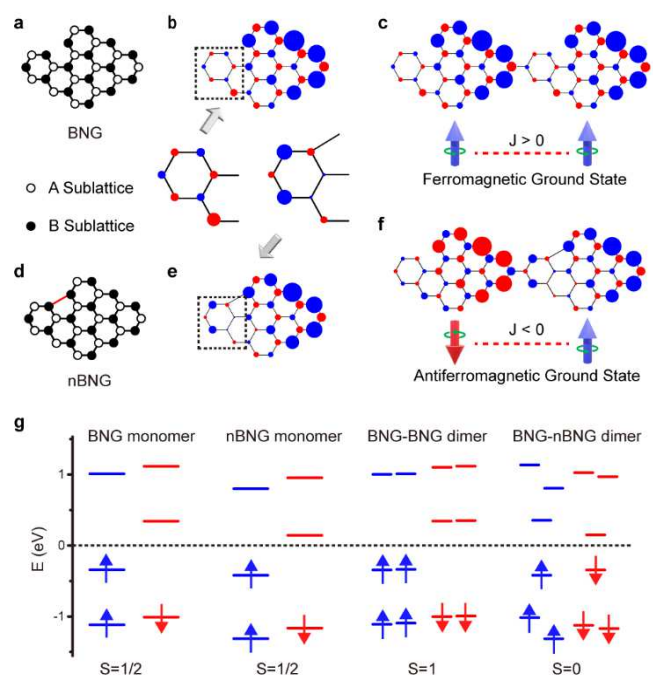

**Fig. 1** Controlling spin order in diradical nanographenes via breaking sublattice symmetry. **a**, Atomic structure of a rhombic bipartite nanographene with a net spin of $S=1/2$ due to sublattice imbalance. **b**, spin density distribution of the nanographene in **a**. **c**, Spin density distribution of a diradical BNG-BNG dimer, exhibiting a ferromagnetic coupled ground state. **d**, Atomic structure of a nonbipartite nanographene with the red C-C bond breaking the bipartite sublattice symmetry of nanographenes. The nonbipartite nanographene also hosts a net spin of $S=1/2$. **e**, spin density distribution of the nanographene in **c**. The presence of five-member ring causes a spin density sign reversal near the red bond, as highlighted by a dashed box. **f**, Spin density distribution of a diradical BNG-nBNG dimer, showing an antiferromagnetic coupled ground state. **g**, Energy level filling of the above four configurations. Blue/red isosurfaces denote spin up/spin down density, respectively.

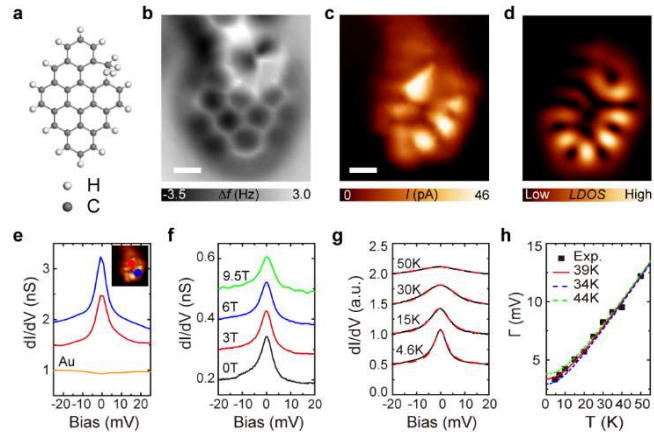

**Fig. 2** Detecting the net spin in a bipartite nanographene monomer via Kondo resonance. **a,** Chemical structure of a rhombic bipartite nanographene. **b,** Nc-AFM frequency shift image (Resonant frequency: 29 KHz, Oscillation amplitude: 160 pm) of the nanographene in **a**. The methyl group is imaged as a bright protrusion due to strong Pauli repulsion. **c-d,** Constant-height current image (Bias voltage: 1 mV) and simulated LDOS map. **e,** d$I$/d$V$ spectra taken on the two locations marked on the inset current image. A sharp zero-energy peak is resolved due to the screening of the net spin of $S=1/2$ in nanographene by Au(111) surface electrons. **f,** Out-of-plane magnetic field dependence of the Kondo resonance. All d$I$/d$V$ spectra were taken at the same position. **g,** Temperature dependence of the Kondo resonance. All d$I$/d$V$ spectra were taken at the same position. The dotted lines are simulated curves using a Frota function. **h,** The full peak width at half maximum as a function of temperature. A Kondo temperature of 39K is obtained. Scale bars: 0.3 nm.

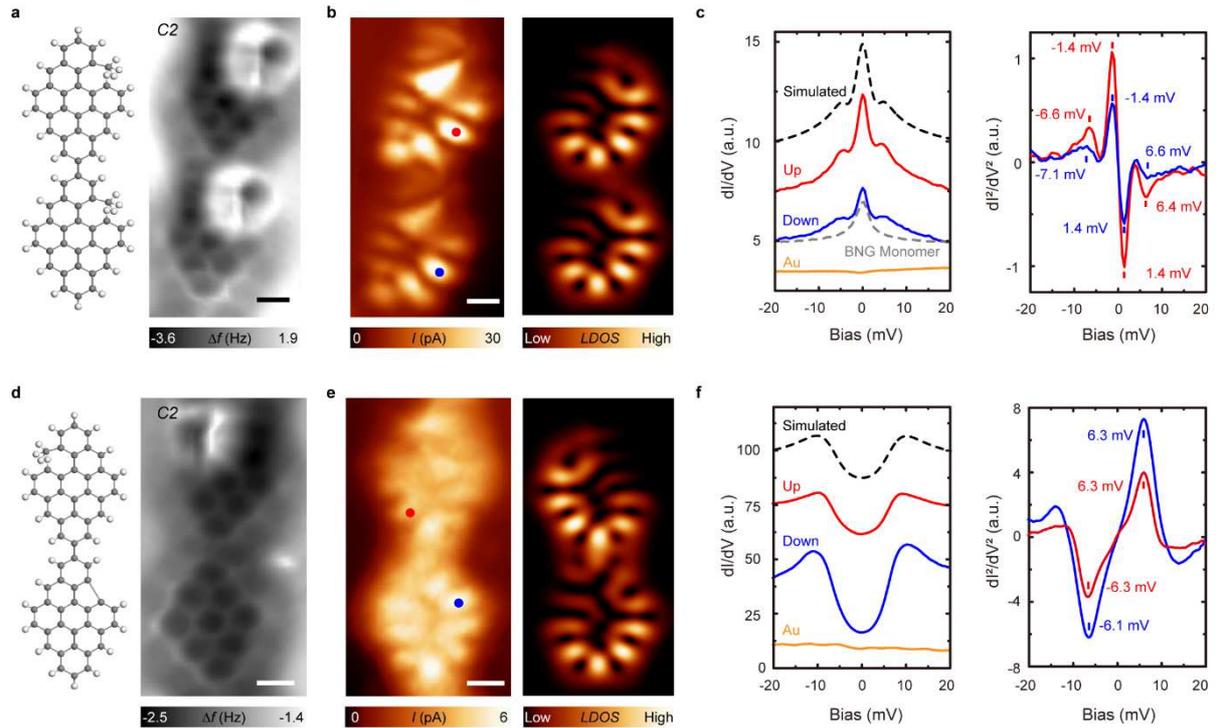

**Fig. 3** Determining the spin order in diradical nanographene dimers by spin-flip spectroscopy. **a**, Chemical structure and nc-AFM image (Resonant frequency: 29 KHz, Oscillation amplitude: 160 pm) of a BNG-BNG dimer. **b**, Constant-height current image (Bias voltage: 1 mV) and simulated LDOS map of the dimer in **a**. **c**, d$I$/d$V$ spectra (modulation: 0.1 mV, Temperature: 1.1K) and numerical calculated d$I^2$/d$V^2$ spectra taken on the two locations marked in **b**. Three resonances have been detected, suggesting a ferromagnetic coupled ground state with $S$=1. The black dashed line is simulated line shape using a perturbative approach up to third order. **d**, Chemical structure and nc-AFM image (Resonant frequency: 22 KHz, Oscillation amplitude: 140 pm) of a BNG-nBNG hetero-dimer. **e**, Constant-height current image (Bias voltage: 1 mV) and simulated LDOS map of the dimer in **d**. **f**, d$I$/d$V$ spectra and numerical calculated d$I^2$/d$V^2$ spectra taken on the two locations marked in **e**. Two symmetric peaks have been detected, suggesting an antiferromagnetic coupled ground state with $S$=0. The black dashed line is simulated line shape using a perturbative approach up to third order. Scale bars: 0.3 nm.

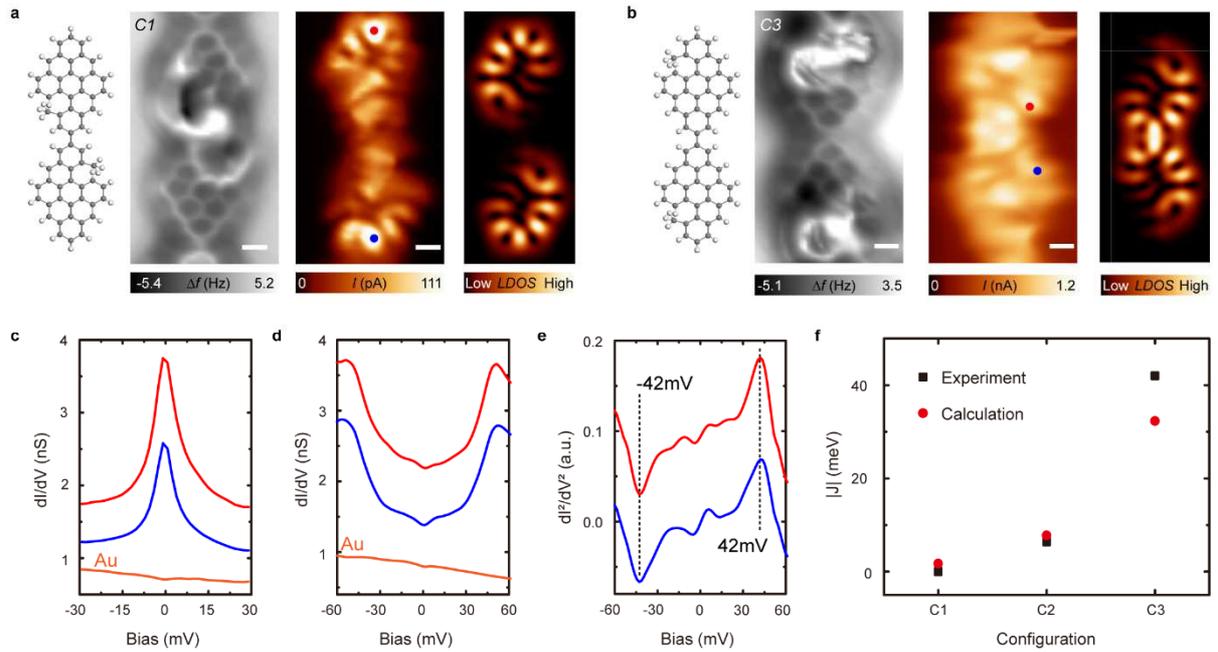

**Fig. 4** Engineering exchange-interaction strength in diradical nanographene dimers. **a**, From left to right: Chemical structure, nc-AFM image (Resonant frequency: 29 KHz, Oscillation amplitude: 160 pm), Constant-height current image (Bias voltage: 1 mV), and simulated LDOS map of a diradical nanographene dimer with minimum spin density overlap of the two spins. **b**, From left to right: Chemical structure, nc-AFM image (Resonant frequency: 29 KHz, Oscillation amplitude: 160 pm), Constant-height current image (Bias voltage: 1 mV), and simulated LDOS map of a diradical nanographene dimer with maximum spin density overlap of the two spins. **c**, d$I$/d$V$ spectra spectra taken on the two locations marked in **a**. The presence of Kondo resonances at each unit in the dimer suggest a negligible exchange interaction. **d-e**, d$I$/d$V$ spectra and numerical calculated d$I^2$/d$V^2$ spectra taken on the two locations marked in **b**. Two symmetric peaks have been detected, suggesting an antiferromagnetic coupled ground state with exchange-interaction strength of 42 meV. **f**, Summarized exchange-interaction strength of three dimer configurations ***C1-C3*** marked in Fig. 3 and Fig. 4. Scale bars: 0.3 nm.

**TABLE 1. Summary the ground state of all dimer configurations.**

| Ground state | Uncoupled spins | S=0 | | S=1 |
|---|---|---|---|---|
| Configuration | *C1* | *C2* | *C3* | *C2* |
| *nBNG-nBNG* | 0 meV (0.15 meV) | 3.4±0.3 meV (3.05 meV) | 35.8±0.4 meV (28.80 meV) | - |
| *BNG-BNG* | 0 meV (0.15 meV) | - | 42.7±0.7 meV (32.30 meV) | -6.6±0.3 meV (-7.75 meV) |
| *BNG-nBNG* | 0 meV (-0.95 meV) | 6.3±0.3 meV (3.20 meV) | 42.9±0.4 meV (30.50 meV) | - |
| *nBNG-BNG* | 0 meV (-0.95 meV) | - | 42.9±0.4 meV (30.50 meV) | -5.9±0.3 meV (-7.35 meV) |

**Exchange-coupling strength *J* extracted from experimental (simulation) data. *C1*, *C2* and *C3* are the connecting configurations denoted in Fig. 3-4 in the main paper.